\documentclass{appolb}

\def\slashchar#1{\setbox0=\hbox{$#1$}
   \dimen0=\wd0 \setbox1=\hbox{/} \dimen1=\wd1
   \ifdim\dimen0>\dimen1 \rlap{\hbox to \dimen0{\hfil/\hfil}} #1
   \else  \rlap{\hbox to \dimen1{\hfil$#1$\hfil}} / \fi}

\usepackage{cite}
\usepackage{epsfig}

\usepackage{graphicx}
\usepackage{subfigure}
\usepackage{cite}

\begin{document}

\title{Gravitational, electromagnetic, and transition form factors of the pion%
\thanks{Presented by WB at {\em MINI-WORKSHOP BLED 2009: PROBLEMS IN MULTI-QUARK STATES}, Bled, Slovenia,  29 June - 6 July 2009}}

\author{Wojciech Broniowski
\address{The H. Niewodnicza\'nski Institute of Nuclear Physics PAN, PL-31342 Krak\'ow\\
and Institute of Physics, Jan Kochanowski University, PL-25406~Kielce, Poland}
\and
Enrique Ruiz Arriola
\address{Departamento de F\'{\i}sica At\'omica, Molecular y Nuclear, Universidad de Granada, E-18071 Granada, Spain}
}

\date{8 October 2009 (ver. 2)}

\maketitle
\begin{abstract}
Results of the Spectral Quark Model for the gravitational,
electromagnetic, and transition form factors of the pion are discussed.
In this model both the parton distribution amplitude and the parton
distribution function are flat, in agreement with the transverse lattice
calculations at low renormalization scales. 
The model predictions for the gravitational form factor are
compared to the lattice data, with good agreement. 
We also find a remarkable relation between the three form factors, holding within our model,
which besides reproducing the anomaly, provides a relation
between radii which is reasonably well fulfilled. Comparison with the
CELLO, CLEO, and BaBar data for the transition form factor is also considered. 
While asymptotically the model goes above the perturbative QCD limit, in qualitative agreement 
with the BaBar data, it fails to accurately reproduce the data at intermediate momenta.
\end{abstract}

 
\bigskip
 
The low-energy behavior of the pion is determined by the spontaneous
breakdown of the chiral symmetry. This fact allows for modeling the
soft matrix elements in a genuinely dynamical way
\cite{Davidson:1994uv,Dorokhov:1998up,Polyakov:1998td,Polyakov:1999gs,Dorokhov:2000gu,Anikin:2000th,Anikin:2000sb,RuizArriola:2001rr,Davidson:2001cc,RuizArriola:2002bp,RuizArriola:2002wr,Praszalowicz:2002ct,Tiburzi:2002kr,Tiburzi:2002tq,Theussl:2002xp,Broniowski:2003rp,Praszalowicz:2003pr,Bzdak:2003qe,Noguera:2005cc,Tiburzi:2005nj,Broniowski:2007fs,Courtoy:2007vy,Courtoy:2008af,Dorokhov:2008pw,Kotko:2008gy}.
This talk is based on Refs.~\cite{Broniowski:2007si,Broniowski:2008hx}
and employs the Spectral Quark Model (SQM) \cite{RuizArriola:2003bs}
in the analysis of several high-energy processes and their partonic
interpretation. This model satisfies {\em a priori} consistency
conditions \cite{RuizArriola:2003bs} between open quark lines and closed quark lines, which
becomes crucial in the analysis of high-energy processes and enables an
unambiguous identification of parton distribution functions and
amplitudes. This is not necessarily the feature of other versions of chiral quark models,
such as the Nambu--Jona--Lasinio (NJL) model, as was spelled out already in 
Ref.~\cite{Davidson:1994uv}. For these reasons SQM is particularly well suited for the 
presented study.

The general theoretical framework is set by the Generalized
Parton Distributions (GPDs)
\cite{Ji:1998pc,Radyushkin:2000uy,Goeke:2001tz,Bakulev:2000eb,Diehl:2003ny,Ji:2004gf,Belitsky:2005qn,Feldmann:2007zz,Boffi:2007yc}. These
objects arise formally, {\em e.g.}, from deeply virtual Compton scattering
(DVCS) on a hadronic target, effectively opening up the quark lines
joining the currents. In local quark models usually the one-loop divergences
appear and a regularization is needed.  One may either compute
the {\em regularized} DVCS and take the high-energy limit, or compute
directly the {\em regularized} GPD. Besides the requirements of gauge invariance and
energy-momentum conservation, this apparently innocuous issue sets a
non-trivial consistency condition on admissible regularizations which
SQM fulfills satisfactorily.

For the case of the pion, the GPD for the non-singlet channel is defined as
\begin{eqnarray}
&& \epsilon_{3ab}\,{\cal H}^{q,{\rm NS}}(x,\zeta,t) \!=\!\! \int
\frac{dz^-}{4\pi} e^{i x p^+ z^-}\!\!\!\! \left . \langle \pi^b (p') | \bar \psi (0)
\gamma^+ \psi (z) \, \tau_3 | \pi^a (p) \rangle \right |_{z^+=0,z^\perp=0}, \nonumber 
\end{eqnarray}
with similar expressions for the singlet quarks and gluons. We omit
the gauge link operators $[0,z]$, absent in the light-cone gauge.  The
kinematics is set by $p'=p+q$, $p^2=p'^2=m_\pi^2$, $q^2=-2p\cdot
q=t$. The variable $\zeta = q^+/p^+$ denotes the momentum fraction
transferred along the light cone.  Formal properties of GPDs can be
elegantly written in the symmetric notation involving the variables
$\xi= \frac{\zeta}{2 - \zeta}$, $X = \frac{x - \zeta/2}{1 - \zeta/2}$:
\begin{eqnarray}
H^{I=0}(X,\xi,t)=-H^{I=0}(-X,\xi,t), \; H^{I=1}(X,\xi,t)=H^{I=1}(-X,\xi,t). \nonumber
\end{eqnarray} 
For $X \ge 0$ one has
${\cal H}^{I=0,1}(X,0,0) = q^{S, NS}(X)$,
where $q(x)^i$ are the standard parton distribution functions
(PDFs). In QCD all these objects are subjected to radiative
corrections, as they carry anomalous dimensions, and become scale-dependent, 
{\em i.e.} they undergo a suitable QCD evolution. This raises an important
question: what is the scale $Q_0$ of the quark model when matching to QCD is performed? 
The momentum-fraction sum rule fixes this scale to be admittedly very low, 
$Q_0 = 313_{-10}^{+20} {\rm ~MeV}$, for $\Lambda_{\rm QCD}=226~{\rm MeV}$. 
Remarkably, but also perhaps unexpectedly, this choice, followed by the leading-order evolution, provides
a rather impressive agreement with the high energy data, as well as the Euclidean and
transverse-lattice simulations (see
Ref.~\cite{Broniowski:2007si} for a detailed summary).

The following {\em sum rules} hold for the moments of the GPDs:
\begin{eqnarray}
\int_{-1}^1 \!\!\!\!\! dX\, {H}^{I=1}(X,\xi,t) = 2 F_V(t), 
\; \int_{-1}^1 \!\!\!\!\! dX\,X \, {H}^{I=0}(X,\xi,t) = 2\theta_2(t)-2\xi^2 \theta_1(t), \nonumber
\end{eqnarray}
where $F_V(t)$ denotes the {\em vector form factor}, while
$\theta_1(t)$ and $\theta_2(t)$ stand for the {\em gravitational form
  factors} \cite{Donoghue:1991qv}.  Other important features are the
{\em polynomiality} conditions~\cite{Ji:1998pc}, the {\em positivity
  bounds}~\cite{Pire:1998nw,Pobylitsa:2001nt}, and a {low-energy
  theorem}~\cite{Polyakov:1998ze}.  We stress that all these
properties required on formal grounds are satisfied in our quark-model
calculation \cite{Broniowski:2007si}. Unlike GPDs, the form factors of
conserved currents do not undergo the QCD evolution.

\begin{figure}[tb]
\begin{center}
\subfigure{\includegraphics[width=.48\textwidth]{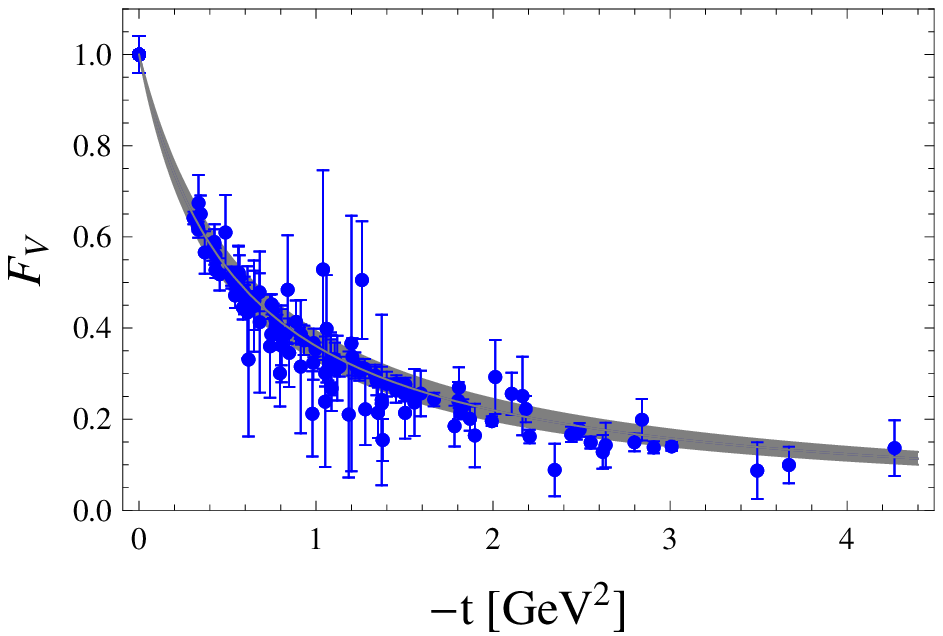}} \hfill
\subfigure{\includegraphics[width=.48\textwidth]{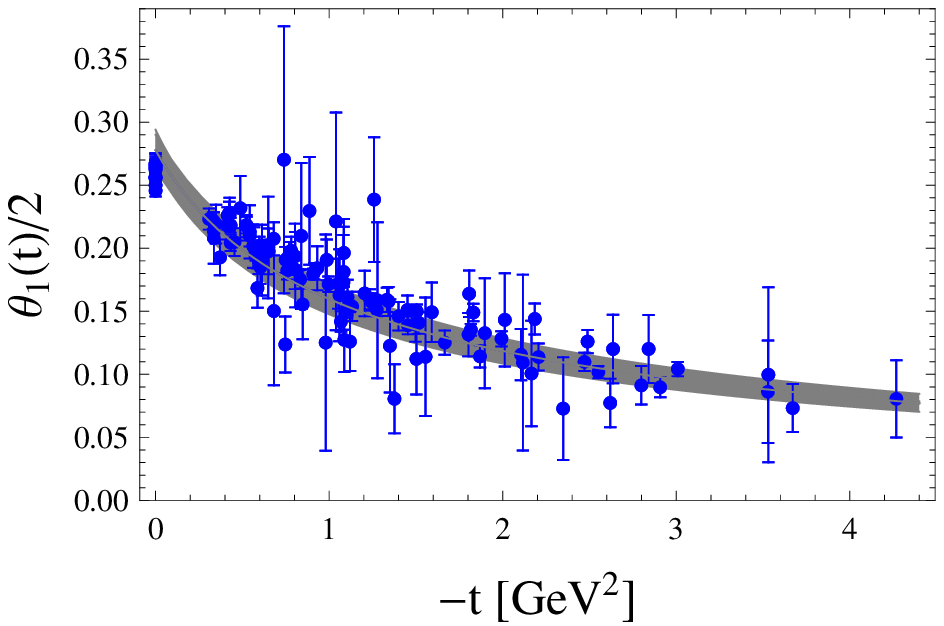}} 
\end{center}
\vspace{-3mm}
\caption{Form factors of the pion vs. lattice data. Left: the electromagnetic form factor. Right: the quark part of the
  gravitational form factor, $\theta_1(t)/2$, computed in the Spectral
  Quark Model and compared to the lattice data from
  Ref.~\cite{Brommel:PhD}. The band around the model curves indicates
  the uncertainty in the model parameters. \label{fig:ff}}
\end{figure}
In the chiral limit we have the following identity in SQM relating the
gravitational and electromagnetic form factor,
\begin{eqnarray}
\frac{d}{dt} \left[ t\, \theta_i (t ) \right] &=& F_V(t ) \, , \quad (i=1,2) \, , 
\label{eq:FV-theta}
\end{eqnarray}
from which the identity between the two gravitational form factors
$\theta_1 (t ) = \theta_2 (t ) \equiv \Theta(t)$ follows.  

Since there
is no data for the full kinematic range for the GPDs of the pion, we
present here the results for the generalized form factors only, in
particular for the gravitational ones. It is well known that the data for
the electromagnetic form factor are well parameterized with the
monopole form, which by construction is reproduced in SQM, where the
vector meson dominance is built in.  The gravitational form factors
are available from the lattice QCD simulations
\cite{Brommel:PhD,Brommel:2005ee}.  In Fig.~\ref{fig:ff} the
electromagnetic form factor and the quark part of the gravitational
form factor are compared to the lattice data.  We note a very good
agreement. In SQM one has the relation
\begin{eqnarray}
m_\rho^2 = 24 \pi^2 f^2/N_c, \label{relmv}
\end{eqnarray}
where $f$ is the pion weak decay constant in the chiral limit. This
relation works within a few percent phenomenologically. The expressions
for the form factors in SQM are very simple,
\begin{eqnarray}
F_V(t)=\frac{m_\rho^2}{m_\rho^2-t}, \;\;\;\; \theta_{1,2}(t)/\theta_{1,2}(0)=\frac{m_\rho^2}{t} \log
\left ( \frac{m_\rho^2}{m_\rho^2-t} \right ). 
\label{eq:fv+theta}
\end{eqnarray}
We note the longer tail of the gravitational form factor in the
momentum space, meaning a more compact distribution of energy-momentum
in the coordinate space. Explicitly, we find a quark-model formula 
\begin{eqnarray}
2\langle r^2 \rangle_\theta = \langle r^2 \rangle_V.
\end{eqnarray}  

\begin{figure}[tb]
\begin{center}
\subfigure{\includegraphics[angle=0,width=0.44\textwidth]{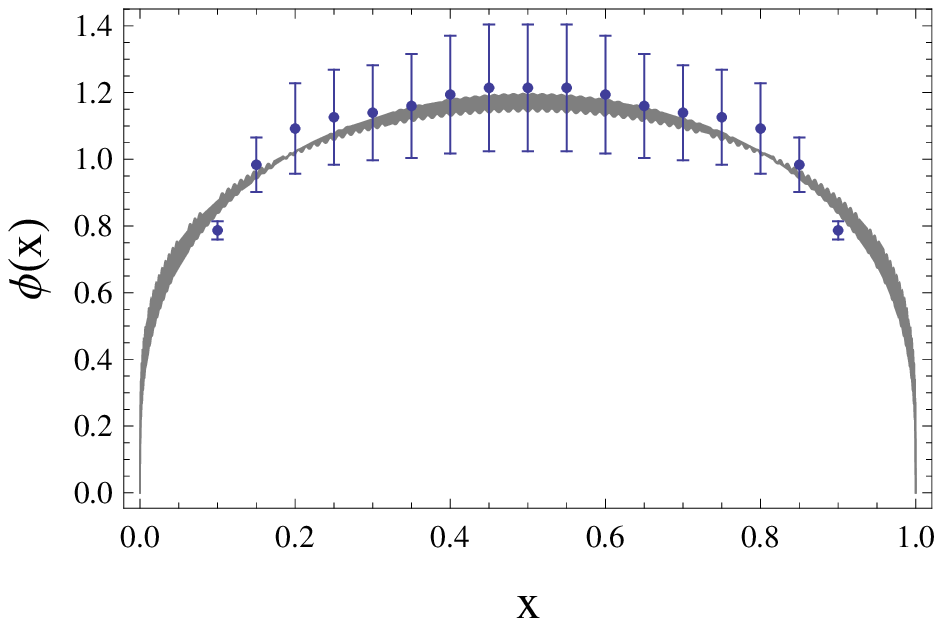}}\hfill
\subfigure{\includegraphics[angle=0,width=0.45\textwidth]{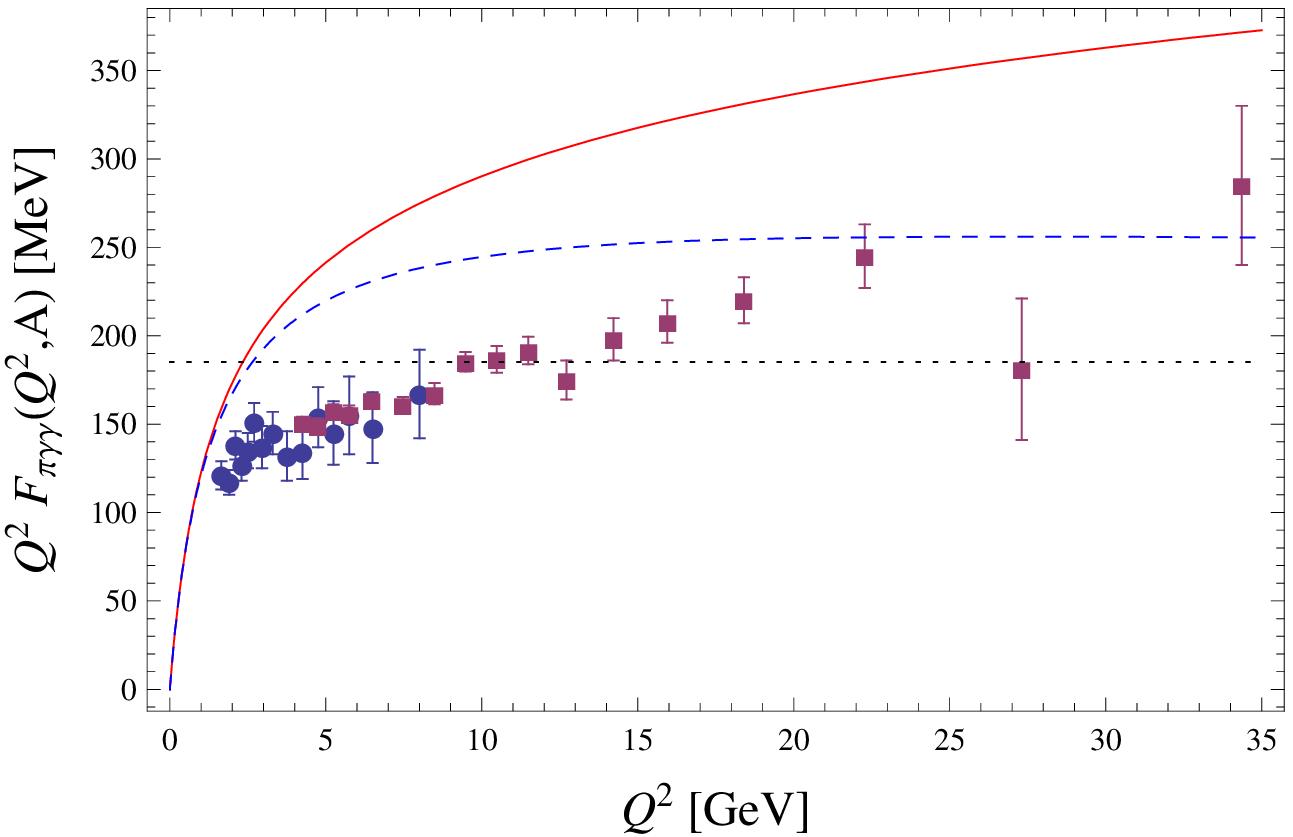}} 
\end{center}
\caption{Left: chiral quark model prediction for the pion DA evolved
  to the scale of $0.5 {\rm GeV}$ (band) and compared to the
  transverse lattice data~\cite{Dalley:2002nj}. Right: the pion
  trantition form factor compared to the CLEO~\cite{Gronberg:1997fj}
  and BaBar~\cite{:2009mc} data.  Solid (dashed) lines are the SQM
  prediction at $A=1$ ($A=0.95$). The dotted line is the perturbative
  QCD prediction.}
\label{fig:pda+babar}
\end{figure}

The two previous processes regard two pions and either one photon or
one graviton in the corresponding three-point vertex function. 
An apparently disparate object is given by the pion-photon 
{\em transition distribution amplitude} (TDA) \cite{Lansberg:2006fv,Lansberg:2007bu}
\begin{eqnarray}
\int \! \frac{d z^-}{2\pi}\, e^{ix p^+ z^-} \!\langle
\gamma(p',\varepsilon)| \bar{\psi}(0)\gamma^\mu
\frac{\tau^a}{2}{\psi}(z) |\pi^b(p) \rangle
\Big|_{\stackrel{z^+=0}{z_T=0}} &=& \!\frac{i e}{p^+ f} \epsilon^{\mu
\nu \alpha \beta}\varepsilon_\nu p_\alpha q_\beta V^{ab}(x,\zeta,t),
\label{VTDA}  
\end{eqnarray}
Here the photon carries momentum $p'=p+q$ and has polarization
$\varepsilon$. As before, the presence of the gauge link operators is
understood in Eq.~(\ref{VTDA}) to guarantee the gauge invariance of the
bilocal operators.
We consider here the {\em isovector} quark
bilinears. Since 
the photon couples to the quark through a combination of the
isoscalar and isovector couplings, {\em i.e.} the quark charge is
$Q=1/(2N_c)+\tau^3/2$, one has the isospin decomposition
\begin{eqnarray}
V^{ab} (x,\zeta,t) &=&\delta^{ab}V_{I=0} (x,\zeta,t) + i \epsilon^{abc}V_{I=1} (x,\zeta,t) .
\label{isodec} 
\end{eqnarray}
The isoscalar form factor is related to the
pion-photon {\it transition form factor} by the sum rule
\begin{eqnarray}
F_{\pi \gamma \gamma^\ast}(t)=\frac{2}{f}\int dx V^{I=0} (x,\zeta,t), 
\end{eqnarray}
where the factor of 2 comes fom the fact, that either of the photons
can be isoscalar. The form factor in SQM was obtained directly in
Ref.~\cite{RuizArriola:2003bs} and later on from the integration of
the pion-photon isoscalar transition distribution amplitude (TDA)
yielding~\cite{Broniowski:2007fs} a $\zeta$-independent function (as required by polynomiality), 
\begin{eqnarray}
F_{\pi \gamma \gamma}(t,A)= \frac{2f}{N_c}\left[ \frac{2 m_\rho^2}{m_\rho^4-t m_\rho^2+(1-A^2)t^2}
+\frac{1}{A t} \log
\left ( \frac{2m_\rho^2-(1-A)t}{2m_\rho^2-(1+A)t} \right ) \right] ,
\label{eq:fpiggA}
\end{eqnarray}
where $A=(q_1^2-q_2^2)/(q_1^2+q_2^2)$ is the photon asymmetry parameter. For $A=1$ we have
\begin{eqnarray}
F_{\pi \gamma \gamma^*}(t)= \frac{1}{12 \pi^2 f }\left[ \frac{2 m_\rho^2}{m_\rho^2-t}+\frac{m_\rho^2}{t} \log
\left ( \frac{m_\rho^2}{m_\rho^2-t} \right ) \right],
\label{eq:fpigg}
\end{eqnarray}
where relation (\ref{relmv}) has been used.  We read out from this formula the
corresponding rms radius to be $\langle r^2 \rangle^{1/2}_{\pi \gamma
  \gamma^\ast}=\sqrt{5}/m_\rho= 0.57 {\rm ~fm}$ for $m_\rho= 770 {\rm ~MeV}$. Equivalently, one may use the slope parameter $b_\pi=
\frac{d}{d t} F_{\pi^0 \gamma \gamma^\ast} (t)/F_{\pi^0 \gamma
  \gamma^\ast} (t)\Big|_{t=0}$.  SQM gives $b_\pi = 5/(6 m_\rho^2) =
1.4~{\rm GeV}^{-2}$, in a very reasonable agreement with the
experimental value $b_\pi = (1.79 \pm 0.14 \pm 14) {\rm GeV}^{-2}$, 
originally reported by CELLO~\cite{Behrend:1990sr}.  A comparison of
Eq.~(\ref{eq:fpiggA},\ref{eq:fpigg}) to the
CLEO~\cite{Gronberg:1997fj} and BaBar~\cite{:2009mc} data is presented
in the right panel of Fig.~\ref{fig:pda+babar}.  The solid line
corresponds to the model calculation with $A=1$, while the dashed line
is for $A=0.95$. We note that the experiment does not produce strictly
real photons, thus the observed sensitivity to the value of $A$ is a
relevant effect. We note that while at $|A|=1$ the model asymptotics
for the transition form factor is $(2f/N_c)\log(-t/m_\rho^2)/(-t)$, at
$|A|\neq 1$ it becomes $(2f/N_c)\log[(1+A)/(1-A)]/(-A t)$. The
behavior is clearly seen in Fig.~\ref{fig:pda+babar}.  As we notice, in the
intermediate range of $Q$ SQM overshoots the data.

The recent BaBar measurements~\cite{:2009mc} have predated the
long-standing perturbative QCD
prediction~\cite{Efremov:1979qk,Lepage:1979zb} that $-t F_{\pi \gamma
  \gamma^\ast}(t)$ goes asymptotically to a constant value of
$2f$. Some authors~\cite{Radyushkin:2009zg,Polyakov:2009je} have
pointed out that the key to this unexpected behavior hints for a flat
pion PDA and the end-point singularities and switched-off QCD
evolution.  The flatness of the PDA at low renormalization scales has
been originally found in the Nambu--Jona-Lasinio
model~\cite{RuizArriola:2002bp} and in SQM~\cite{RuizArriola:2003bs}.

We note in passing that a constant PDA is also found in the Regge model \cite{RuizArriola:2006ii}.

Remarkably, an almost flat PDA is also found non-perturbatively on the
transverse lattice~\cite{Dalley:2002nj} (see the left panel of
Fig.~\ref{fig:pda+babar}). Actually, the non-vanishing of the PDA at the
end points (at the quark-model scale) is not only a consequence of
local quark models. Nonlocal models correctly implementing the chiral
Ward-Takahashi identity also get such a feature~\cite{Bzdak:2003qe}. A
trend to flatness is observed in contrast to calculations violating
the chiral symmetry constraints. However, the corresponding transition
form factor in non-local models does not show a steep
rise~\cite{Kotko:2009ij} as suggested by the BaBar data.  The
calculation in Ref.~\cite{Dorokhov:2009dg,Dorokhov:2009zx}, which reproduces the CLEO
and BaBar data, requires, unfortunately, a much too small constituent
quark mass, which is incompatible with other sectors of the pion
phenomenology. 
The apparent inconsistency of the BaBar data with the QCD convolution scheme is 
also addressed in Ref.~\cite{Mikhailov:2009kf,Mikhailov:2009ah}.

Let us remind the reader that according to the conventional
perturbative QCD approach, the radiative corrections are computed
order by order in the twist expansion. Most often this is in practice
possible only for the leading-twist contribution. Actually, this is
the only way to identify the PDA within a non-perturbative scenario or
quark model calculations. In fact, the chiral quark models require a low
scale not only by fixing the second Gegenbauer coefficient $a_2$ of
the PDA. As already mentioned, the same conclusion is reached independently by fixing the
momentum fraction of the valence quarks to its natural $100\%$ value
at the quark-model scale, where the quarks constitute the only degrees
of freedom.

On a more methodological level, it is worth mentioning that the
conventional NJL model does not share some of the virtues of SQM,
particularly the interplay between chiral anomaly and factorization, a
subtle point which was discussed at length in
Ref.~\cite{RuizArriola:2002wr} for the NJL case. The $\pi \gamma
\gamma$ triangle graph is linearly divergent, and thus a
regularization must generally be introduced. If one insists on
preserving the vector gauge invariance, the regulator must preserve that
symmetry, but then the axial current is not conserved, generating the
standard chiral anomaly. The obvious question arises whether the limit
$Q^2 \to \infty $ must be taken before or after removing the cut-off.
If one takes the sequence $Q^2 > \Lambda^2$, a constant PDA is obtained in agreement with our low
energy calculation. For the opposite sequence factorization does not
hold in NJL. The good feature of SQM is that the spectral
regularization does not make any difference between the two ways. 
This illustrates in a particular case the above-mentioned general consistency requirement between regularized open and
closed quark lines (see e.g.~\cite{Arriola:2006ds}).

Finally, by combining Eq.~(\ref{eq:fv+theta}) and
Eq.~(\ref{eq:fpigg}) we get the remarkable relation among 
the electromagnetic, gravitational and transition
form factors, holding in SQM:
\begin{eqnarray}
F_{\pi \gamma \gamma^*} (t)= \frac{1}{12 \pi^2 f }\left[ 2
  F_V(t)+\Theta(t) \right]\, , 
\label{eq:3ff}
\end{eqnarray}
whence 
\begin{eqnarray}
3 \langle r^2 \rangle_{\pi \gamma \gamma^*} = 2 \langle r^2
\rangle_V + \langle r^2 \rangle_\Theta \, . 
\end{eqnarray}
The previous relation is not
fulfilled in the conventional NJL model. Of course, it would be
interesting to test the relation Eq.~(\ref{eq:3ff}) against the future
data or lattice QCD.

In conclusion, we note that while the description of the pion
transition form factor in a genuinely dynamical way remains a
challenge, the Spectral Quark Model offers many attractive features
which are required from theoretical consistency. It satisfies the
chiral anomaly and the factorization property. The vector and
gravitational form factors describe experimental and/or lattice-QCD
data satisfactorily. A remarkable model relation among the
gravitational, electromagnetic and transition form factors has also
been deduced. Finally, for the latter, we have also displayed a
hitherto unnoticed sensitivity to the photon momentum asymmetry parameter $A$ which
might be relevant for other studies.

\bigskip
This research is supported in part by the Polish Ministry of Science and Higher
Education, grants N202~034~32/0918 and N202~249235, Spanish DGI and
FEDER funds with grant FIS2008-01143/FIS, Junta de Andaluc{\'\i}a
grant FQM225-05, and the EU Integrated Infrastructure Initiative
Hadron Physics Project, contract RII3-CT-2004-506078.


\end{document}